\documentclass[11pt,a4paper]{article}

\usepackage[utf8]{inputenc}
\usepackage[T1]{fontenc}
\usepackage[english]{babel}
\usepackage{geometry}
\usepackage{amsmath, amssymb, amsfonts}

\usepackage{booktabs}
\usepackage{array}
\usepackage{tabularx}
\usepackage{longtable}
\usepackage{multirow}

\usepackage{graphicx}

\usepackage{hyperref}
\hypersetup{
    colorlinks=true,
    linkcolor=blue,
    citecolor=blue,
    urlcolor=blue
}

\usepackage[numbers]{natbib}

\usepackage{microtype}
\usepackage{parskip}

\newcolumntype{L}[1]{>{\raggedright\arraybackslash}p{#1}}
\newcolumntype{C}[1]{>{\centering\arraybackslash}p{#1}}

\title{\textbf{Recursivism: An Artistic Paradigm for Self-Transforming Art in the Age of AI}}

\author{
Florentin Koch\\
\'Ecole Polytechnique --- HEC Paris\\
\texttt{florentin.koch@polytechnique.edu}
}

\date{} 

\begin{document}

\maketitle

\begin{center}
\textit{Preprint --- under review.}
\end{center}

\begin{abstract}
This article introduces \textbf{Recursivism} as a conceptual framework for analyzing contemporary artistic practices in the age of artificial intelligence. While recursion is precisely defined in mathematics and computer science, it has not previously been formalized as an aesthetic paradigm. Recursivism designates practices in which not only outputs vary over time, but where the generative process itself becomes capable of reflexive modification through its own effects.

The paper develops a five-level analytical scale distinguishing simple iteration, cumulative iteration, parametric recursion, reflexive recursion, and meta-recursion. This scale clarifies the threshold at which a system shifts from variation within a fixed rule to genuine self-modification of the rule itself. Building on this framework, art history is reinterpreted as a recursive dynamic alternating between internal recursion within movements and meta-recursive transformations of their generative principles.

Artificial intelligence renders this logic technically explicit. Learning loops, parameter updates, and code-level self-modification literalize recursive structures that had previously remained implicit in artistic processes. To distinguish Recursivism from adjacent notions such as generative art, cybernetics, process art, and evolutionary art, the article proposes three operational criteria: state memory ($\mu$), rule evolvability ($\rho$), and reflexive visibility ($R$).

These concepts are examined through detailed case studies, including Refik Anadol's immersive installations, Sougwen Chung's human--machine co-drawing systems, Karl Sims's \textit{Genetic Images}, and the Darwin--Gödel Machine as a contemporary example of constrained meta-recursion in code. The article concludes by examining the aesthetic, curatorial, and ethical implications of self-modifying artistic systems, positioning Recursivism as a conceptual response to the automation of execution and the rise of recursive AI architectures.
\end{abstract}


\vspace{1em}
\hrule
\vspace{1em}

\section{Scope and Contribution}

Each major technical mutation has reconfigured art: the printing press disseminated Renaissance perspective \cite{ivins1953}, photography catalyzed Impressionism \cite{scharf1968}, digital computing opened algorithmic art \cite{paul2015}. Artificial intelligence constitutes a new threshold calling for a theoretical framework that transcends stylistic labels to grasp the logic of auto-transformation in contemporary practices.

While recursion has precise definitions in mathematics and computer science \cite{graham1994} (illustrated by the Fibonacci rule $F(n) = F(n-1) + F(n-2)$, where the same function applies itself to smaller subproblems), the term has never been used to designate a structured artistic movement. Existing references to ``recursive art'' concern only visual motifs such as Droste-like self-similarity \cite{hofstadter1979}.

We introduce \textit{Recursivism} as a neologism for practices in which productions unfold iteratively, the process reconfigures itself through its own effects, and the resulting dynamic of self-transformation becomes a reflexive component of the work. The terminological choice also follows by exclusion: \textit{Iterativism} would imply mere repetition without structural change; \textit{Recurrentism} remains tied to fixed-rule mathematical sequences; and expressions such as \textit{autotransforming art} lack conceptual scope and formal anchoring. The three points below provide the positive justification for adopting the term Recursivism.

\begin{enumerate}
    \item It generalizes the core logic of recursion: reapplying a rule to the results of its previous steps, by extending it to evolving generative rules.
    \item It aligns with contemporary AI systems, whose learning loops, parameter updates, and internal rewritings technically realize self-transforming recursive processes, rather than externally executed iterations.
    \item It preserves the mathematical lineage required for the quasi-formal framework developed later (memory $\mu$, evolvability $\rho$, reflexivity $R$), which aims to describe artistic processes through the logic of dynamic systems.
\end{enumerate}

This article proposes nine parts: (1)~formalize Recursivism's levels of self-modification through a five-level scale; (2)~show how art history alternates between internal recursion and meta-recursion; (3)~propose AI as catalyst of a transition toward conceptual creativity; (4)~show how Recursivism responds to art's dual impasse; (5)~distinguish recursive works through operational criteria; (6)~explain how AI renders recursion technically explicit; (7)~analyze emblematic case studies; (8)~examine aesthetic and curatorial implications; (9)~address critical perspectives and limitations.

\section{Scale from Iteration to Meta-Recursion}

Any act of creation involves two dimensions: a product and a process. To analyse artistic processes with precision, it is necessary to distinguish practices in which only the production varies from those in which the process itself transforms. Across multiple domains, one finds systems whose behaviour depends on their own history: in biological evolution \cite{darwin1859}, where inherited variations shape future states without altering the evolutionary mechanism itself; and in cognitive or living systems \cite{maturana1980}, which adjust their functioning through internal feedback. These examples show that dynamic systems are guided by their past.

To clarify these distinctions, we introduce a scale ranging from simple iteration to full meta-recursion. Each level specifies what changes from one step to the next: state, parameters, rules, or meta-rules; and thus marks the threshold at which a procedure moves from being merely iterative (variation of the product under a constant process) to genuinely self-modifying (variation of the process itself), in accordance with the conceptual framework of Recursivism.

\newpage

\begin{table}[htbp]
\centering
\caption{Recursivity Scale in Artistic Practices}
\label{tab:recursivity-scale}
\small
\begin{tabularx}{\textwidth}{c L{2cm} L{3.5cm} L{2cm} L{2.5cm} L{3cm}}
\toprule
\textbf{Level} & \textbf{Label} & \textbf{Formal expression} & \textbf{Variable element} & \textbf{Reflexivity type} & \textbf{Artistic example} \\
\midrule
0 & Simple iteration & $O_{n+1} = f(O_n)$ & Input changes; rule fixed & None (fixed rule, no memory) & Repeating the same filter on each input \\
\addlinespace
1 & Cumulative iteration & $O_{n+1} = f(O_0, \ldots, O_n)$ & Aggregated inputs & External memory (additive accumulation) & Layering outputs without changing the rule \\
\addlinespace
2 & Parametric recursion & $\begin{cases} O_{n+1} = f(O_n; p_n) \\ p_{n+1} = g(O_n, p_n) \end{cases}$ & Parameters vary & Adaptive feedback (self-adjusting parameters) & AI model updating its weights \\
\addlinespace
3 & Reflexive recursion & $\begin{cases} O_{n+1} = f_n(O_n) \\ f_{n+1} = F(f_n, O_n) \end{cases}$ & Generative rule evolves & Structural self-modulation (rule rewriting under stable metarule $F$) & System modifying its own algorithm \\
\addlinespace
4 & Meta-recursion &$\displaystyle (f_{n+1}, O_{n+1}) = f_n\bigl((f_n, O_n)\bigr)$ & Generative principle evolves & Full self-organization & Redefining the logic governing its own evolution \\
\bottomrule
\end{tabularx}
\end{table}

\noindent\textbf{Legend:} $O_n$: output at iteration $n$; $p_n$: internal parameters; $f$, $g$, $F$: fixed functions; $f_n$: evolving function.

\noindent\textbf{Interpretation:} Levels 0--1 vary outputs; Level 2 varies parameters; Level 3 varies generative rules; Level 4 varies metarules.

\section{Transition: From Formal Framework to Historical Dynamics}

Viewed through the preceding framework, art history can be reinterpreted as a recursive dynamic with two layers: internal recursion, where a movement iterates and refines its own rule, and meta-recursion, where that rule itself becomes the object of transformation. The descriptive task of this section is to show how this logic structures major historical shifts; the prescriptive task of subsequent sections is to argue that AI now operates as a catalyst for a new artistic paradigm, one capable of addressing the longstanding contradictions revealed by the trajectory of art history.

\section{Recursion in Art History: Toward Recursivism}

Art history functions as a two-level recursive system: each movement repeats and transforms the rules that found it.

\subsection{Internal Recursion}
Within a movement, the same rule applies to its own productions. Realism perfects mimesis, Impressionism refines light, the avant-garde explores medium deconstruction. Each work extends the previous according to a stable protocol (forms of recursivity 0, 1 on our scale).

\subsection{Meta-Recursion}
When this iteration exhausts itself, the rule itself becomes object of transformation. Minimalism does not perfect figuration; it questions the very possibility of representation. Conceptual art does not refine technique; it investigates what makes an act ``artistic.'' These are not new styles but meta-operations: movements that act on the generative principle of movements.

This dialectical dynamic characterizes art's evolution: internal recursion (optimizing within a paradigm) until saturation, then meta-recursion (paradigm transformation). Each major rupture (Renaissance, Modernism, Conceptualism) operates this meta-recursive jump.

\section{AI as Historical Operator: The Impressionism Analogy}

The mechanical reproduction enabled by photography provoked an existential crisis for painting. Baudelaire called it ``the refuge of all failed painters'' and ``art's most mortal enemy'' \cite{baudelaire1965}. Yet Impressionism (1860--1880) turned that threat into freedom, abandoning imitation for perceptual immediacy, color vibration, and gesture; dimensions photography could not capture.

Artificial intelligence now revives a similar tension. As it perfects technical execution, the artist's manual role recedes, but this shift opens space for conceptual authorship: designing recursive systems, feedbacks, and data ecologies. Art moves from making to architecting processes.

Recursivism, in this sense, marks not theory but historical necessity. The analogy with photography remains heuristic, not demonstrative, yet the pattern persists: each technology that automates execution forces art to redefine itself through reflection on its own production.

\section{Recursivism Facing the Impossibility of Foundation}

Modern art, seeking to reduce the artwork to its essence, engaged in two complementary paths revealing the same impossibility: fixing art once and for all.

\subsection{Impasse of Stylistic Constancy}
Each movement sought definitive style: realism's mimesis, cubism's geometry, abstraction's purity. But every style engenders its negation. The proliferation of forms shows none embodies art's truth; what persists is the dynamic of supersession itself.

\subsection{Impasse of Ultimate Minimalist Foundation}
Malevich with \textit{Black Square} (1915) \cite{malevich1990} and Reinhardt with \textit{Ultimate Paintings} (1960--1967) \cite{reinhardt1991} each believed they touched painting's ontological ground. But every proclaimed foundation erases itself: each ``last painting'' calls for another, revealing that absolute void does not exist.

However, Derrida's ``logic of the supplement'' \cite{derrida1976} shows external addition is already interior to structure. Thus, neither style nor foundation can serve as stable anchor; what remains is the constancy of the process that self-modifies while creating.

\subsection{Recursivism's Response: Process Constancy}
Faced with these twin impasses, Recursivism formulates a third way: stability resides not in products but in the constancy of the transformative process itself.

Rather than seeking what art \textit{is}, Recursivism focuses on how art \textit{becomes}. The constant is not the result but the recursive operation: the capacity of a system to act on itself and reconfigure based on its own effects.

This shift has profound implications: the artwork is no longer a fixed object but a documented process trajectory; aesthetic value transfers from formal excellence to systemic coherence of the recursive loop; the artist becomes architect of transformation rules rather than producer of finished forms.

\section{Recursivism vs Related Concepts: Criteria of Distinction}

\subsection{Operational Framework: What is Recursivism}

Recursivism is not only an aesthetic heuristic but a structured analytical model. It can be examined empirically through three core properties of recursive artistic processes: memory, rule evolution, and reflexive visibility. These properties provide a basis for distinguishing recursive practices from other ones.

\subsubsection{Memory ($\mu$): Continuity Across States}

\textbf{Qualitative definition.} Memory refers to how much a process retains from its previous states; whether stylistic traces, structural motifs, behavioural tendencies, parameter histories, or conceptual residues.

\textbf{Symbolic intuition.} Successive outputs $O_n$ and $O_{n+1}$ share information to the degree that they remain recognisably connected:
\begin{equation}
\mu \approx \text{sim}(O_n, O_{n+1})
\end{equation}
where $\text{sim}$ stands for any perceptual or structural similarity relation.

\textbf{Interpretation.} High $\mu$ indicates continuity and inheritance; low $\mu$ signals rupture or drift.

\subsubsection{Evolvability ($\rho$): Transformation of the Generative Rule}

\textbf{Qualitative definition.} Evolvability captures how the generative rule: the mapping that produces each new state; modifies itself over time. This modification may involve code rewriting, algorithmic adjustment, shifting protocols, or conceptual reframing.

\textbf{Symbolic intuition.}
\begin{equation}
\rho \approx \Delta\,\text{rule}_n
\end{equation}
where $\Delta$ denotes any effective change detectable in the rule as the process unfolds.

\textbf{Interpretation.} High $\rho$ occurs when the system rewrites its own logic; low $\rho$ corresponds to stable, rule-driven generation.

\subsubsection{Reflexivity ($R$): Visibility of Process Structure}

\textbf{Qualitative definition.} Reflexivity concerns how much the structure of the process manifests in its outputs: loops, recombinations, depth of iteration, self-evaluation, or meta-patterns.

\textbf{Symbolic intuition.}
\begin{equation}
R \approx \text{obs}(\text{rule} \to O)
\end{equation}
meaning the extent to which observing the outputs allows an informed viewer to infer the generative dynamics.

\textbf{Interpretation.} High $R$ indicates transparent or self-revealing processes; low $R$ indicates opaque or black-box dynamics.

\subsubsection{Integrated Conceptual Space}

Taken together, $\mu$, $\rho$, and $R$ form a triangular operational space describing how recursive systems behave. They serve as diagnostic lenses for identifying recursive processes, comparative tools for mapping how different artistic or hybrid systems evolve, and a common vocabulary linking human, machine, and human--machine recursivism practices.

\subsubsection{Methodological Horizon}

Although $\mu$, $\rho$ and $R$ are introduced here as conceptual descriptors, several methodological avenues could support their empirical approximation in future work. Information theory \cite{shannon1948} offers abstract tools (mutual information, compressibility), but algorithmic similarities rarely align perfectly with perceptual or aesthetic similarity, especially across media. Embedding-based approaches \cite{russell2021} (latent spaces for images, gestures, sounds, or text) provide geometric distances that can approximate perceived continuity, yet they depend heavily on the chosen representation and training data. Such developments, however, would require a dedicated methodological study beyond the scope of this article.

\subsection{What is Not Recursivism}

Across the history of art and computation, numerous practices have explored iteration, feedback, emergence, or self-reference. However, a structural analysis reveals fundamental differences in their operational logic.

Generative Art \cite{galanter2003, boden2009} typically relies on combinatorial exhaustion within fixed constraints; the primary goal is variation within a pre-defined search space. First-order Cybernetics \cite{wiener1948} introduces feedback to maintain equilibrium (homeostasis) without altering governing laws. Even Second-order Cybernetics and Autopoiesis \cite{vonfoerster1974, maturana1980}, while foregrounding reflexivity and organizational closure, primarily describe systems striving to maintain their identity against environmental perturbations (conservation). Process Art \cite{morris1969} privileges making over outcome, and Interactive Systems \cite{cornock1973} emphasize real-time responsiveness. Finally, Evolutionary Art \cite{whitelaw2004} introduces selection and mutation, yet typically optimizes parameters under a fixed fitness function rather than rewriting the evolutionary logic itself.

These approaches form a broad family of dynamic systems that may look dynamic or adaptive, yet share a common structural limitation regarding agency: the generative rule remains constant. Their behaviors unfold \textit{within} a rule, not \textit{on} the rule.

Recursivism reframes this landscape by defining reflexivity not merely as observation, regulation, or optimization, but as \textit{ontological evolution}, namely the capacity of a system to act upon and transform the generative procedures that govern it.

\begin{table}[htbp]
\centering
\caption{Comparative Taxonomy of Creative Systems}
\label{tab:comparative-taxonomy}
\small
\begin{tabularx}{\textwidth}{L{2.2cm} C{1.3cm} C{1.3cm} L{2.2cm} L{2.2cm} C{1.5cm}}
\toprule
\textbf{Framework} & \textbf{Feedback} & \textbf{State Memory ($\mu$)} & \textbf{Rule Evolution ($\rho$)} & \textbf{Dominant Teleology} & \textbf{Recursivism Level} \\
\midrule
Generative Art & No & No & No & Variation (Combinatorics) & 0 \\
\addlinespace
Cybernetics & Yes & Optional & Regulatory (Homeostasis) & Equilibrium (Stability) & 0--1 \\
\addlinespace
Process Art & No & Implicit & No & Action (The making) & 0 \\
\addlinespace
Interactive & Yes & No & No & Responsiveness & 0--1 \\
\addlinespace
Evolutionary Art & Yes & Yes & Selection (Parametric) & Optimization (Fitness) & 2 \\
\addlinespace
\textbf{Recursivism} & Yes & Yes & Structural (Self-rewriting) & Transformation (Becoming) & 2--4 \\
\bottomrule
\end{tabularx}
\end{table}

\newpage

\section{AI as Recursive Catalyst}

Artificial intelligence does not invent recursivism in art; it renders it literal and scalable. Machine learning systems operate through iterative self-modifying loops: training on data, adjusting weights, producing outputs that inform next training cycles \cite{russell2021}.

AI makes three decisive contributions:
\begin{enumerate}
    \item \textbf{Automation} of the recursivism process;
    \item \textbf{Scalability} revealing emergent patterns invisible at human scales;
    \item \textbf{Explicit formalization} as readable code.
\end{enumerate}

However, this technical capability does not automatically produce recursivism. Many AI artworks remain at level 0--1: they generate variations without systemic self-modification. Recursivism requires deliberate architectural design where the artist constructs the feedback topology.

\section{Case Studies: Four Modalities of Contemporary Recursivism}

These case studies illustrate the four recursive levels as operational categories, selected to span the full spectrum of recursive depth.

\subsection{Level 0--1: Cumulative Iteration, Variation Without Feedback (Refik Anadol)}

Anadol's immersive installations \cite{anadol2021} occupy the boundary between Level 0 and Level 1, revealing the importance of distinguishing technical architecture from phenomenological experience.

\textbf{Technical dimension (Level 0):} The GAN models are pre-trained and fixed. Once deployed, no further learning occurs: each image is generated independently from random latent vectors ($O_{n+1} = f(z_{n+1})$), with constant rule and independent inputs. The process includes neither state memory nor feedback from outputs: pure simple iteration.

\textbf{Phenomenological dimension (Level 1 appearance):} Two factors create perceptual continuity: (1) audiovisual montage sequences projections simulating evolving memory; (2) some installations incorporate real-time data streams (meteorological sensors, visitor movements) modulating generation parameters, introducing weak cumulative iteration (external inputs aggregated without rule transformation). The continuous visual flow gives viewers an impression of a system that remembers and evolves, though this is aesthetic staging rather than recursive computation.

\textbf{Operational verdict:} By the three criteria ($\mu$, $\rho$, $R$), Anadol's core architecture remains non-recursive at Level 0. Specific data-driven installations approach Level 1 through parameter modulation, but never achieve genuine feedback loops where outputs reshape the generative rule.

\begin{quote}
\textit{Figure 1. [Image removed in preprint version]}\\
Work referenced: Refik Anadol, \textit{Machine Hallucinations: Nature Dreams}, 2021. Immersive data-sculpture projection, generative adversarial network (GAN), real-time visualization, dimensions variable. Installation view at König Galerie, Berlin, Germany. \copyright\ Refik Anadol Studio.
\end{quote}

\subsection{Level 2: Parametric Recursion (Sougwen Chung)}

Sougwen Chung's \textit{Drawing Operations} series (2015--present) exemplifies parametric recursion. The artist draws collaboratively with robotic arms (D.O.U.G. system) trained on her previous drawing gestures \cite{chung2018}.

The recursive architecture operates as follows:
\begin{enumerate}
    \item \textbf{Initial training:} The robot learns from Chung's gestural database.
    \item \textbf{Co-creation:} Human and robot draw simultaneously; robot responds to human gestures in real-time.
    \item \textbf{Data accumulation:} Each session generates new gestural data.
    \item \textbf{Retraining:} Robot's model is periodically updated with accumulated data.
    \item \textbf{Style drift:} Over iterations, the robot's ``style'' evolves, influencing Chung's subsequent gestures.
\end{enumerate}

This creates a genuine feedback loop: the robot learns from the artist, the artist responds to the robot's interpretations, and this interaction generates data that reshapes future robot behavior. The system manifests state memory (cumulative gesture database) and readable recursive rule (the training-response-retraining cycle).

Chung's work reveals a key Recursivism principle: the artist designs not the drawing but the co-evolutionary system. The aesthetic object is not the final image but the documented trajectory of mutual adaptation.

\begin{quote}
\textit{Figure 2. [Image removed in preprint version]}\\
Work referenced: Sougwen Chung, \textit{Drawing Operations Unit: Generation 2 (D.O.U.G.\_2)}, 2017. Human--robot collaborative drawing installation, recurrent neural network trained on the artist's gestures, robotic arm (UR5), custom software, ink on paper, dimensions variable. \copyright\ Sougwen Chung.
\end{quote}

\subsection{Levels 3 and 4: Reflexive and Meta-Recursion}

Evolutionary algorithms provide an exemplary framework for examining higher orders of recursion in computational and artistic systems. Their architecture: variation, evaluation, selection; remains structurally simple, yet the depth of recursion depends on what evolves: data, generative functions, or the very principles of learning.

\subsubsection{From Iteration to Reflexive Recursion (Level 3)}

Classical Darwinian evolution applies a fixed rule: states change, the rule remains (pure iteration). Genetic programming (Koza 1992) introduced a crucial shift: the individuals in a population are not static data but executable programs, often expressed as trees of mathematical operations. Mutations modify syntax ($+ \to *$, or wrapping a branch in $\sin(\cdot)$), and crossover exchanges subtrees between parent programs. Each individual represents a generative function that produces an image, sound, or structure when executed.

The evolutionary engine stays constant but acts upon rules; the system reprograms itself through its own outputs. This defines reflexive recursion: Level 3.

\subsubsection{Genetic Images (Karl Sims, 1991--1993): Evolution of Generative Rules}

Presented at the Centre Pompidou and Ars Electronica, \textit{Genetic Images} by Karl Sims \cite{sims1991} translates Darwinian evolution into an interactive visual environment. A Connection Machine generates sixteen evolving images, each corresponding to a mathematical function such as:
\begin{equation}
\text{image}(x,y) = \sin(x \cdot y) + \cos(\text{noise}(x))
\end{equation}

Viewers stand on floor sensors before the images they find aesthetically appealing. The corresponding equations are selected, crossed, and mutated to produce the next generation. Over iterations, new terms and structural mutations yield increasingly complex and unpredictable visual behaviors.

Evolution now acts on the level of the code itself: individuals are generative functions rather than static outputs; memory is cumulative (partial inheritance of equations); evaluation is external (human aesthetic choice). The artwork evolves not through changing images but through the transformation of its own generative grammar. This is Level 3: reflexive recursion, or structural self-modulation.

\begin{quote}
\textit{Figure 3. [Image removed in preprint version]}\\
Work referenced: Karl Sims, \textit{Genetic Images}, 1991--1993. Interactive installation using evolutionary algorithms, Connection Machine CM-2 (32,768 processors), sixteen CRT monitors, floor sensors, evolutionary software (LISP). Exhibited at Centre Georges Pompidou (Paris) and Ars Electronica (Linz). \copyright\ Karl Sims / Thinking Machines Corporation.
\end{quote}

\subsubsection{Toward Meta-Recursion (Level 4): Evolution of Learning Principles}

Level 4 designates the maximal conceivable form of recursion: a system capable of rewriting not only its generative rules but also the meta-principle that governs how these rules are evaluated. This upper bound can be expressed as:
\begin{equation}
(f_{n+1},\, O_{n+1}) = f_n\bigl(f_n,\, O_n)
\end{equation}
where the generative function takes itself as an argument and produces its own successor.

In existing systems, one encounters a more constrained form:
\begin{equation}
f_{n+1} = F(f_n, O_n)
\end{equation}
in which the evaluation function evolves, but only under fixed higher-order functions. In \textit{Genetic Images}, evaluation remains external and invariant (human judgment). A fully meta-recursive system would internalize this dimension, autonomously redefining what counts as ``success.''

To date, no artificial or biological system implements the unconstrained form of Level 4. Even natural evolution, capable of reshaping developmental programs and regimes of variation, operates under a fixed evaluative principle: differential selection.

\subsubsection{The Darwin--Gödel Machine: Formal Meta-Recursion}

The Darwin--Gödel Machine \cite{zhang2025} proposes a pragmatic instantiation of Level 4 in the domain of code. Its architecture combines: a self-referential code agent capable of reading and modifying its own Python codebase; and a Darwinian, open-ended exploration loop that generates self-modifications, evaluates them on tasks, and retains those that improve performance.

Exploration generates variants; empirical evaluation selects which become the new agent version. The system can thus rewrite its problem-solving routines, its self-improvement procedures, and part of its own evolutionary mechanism. The environment and global objective (benchmark performance) remain fixed, but the internal logic of adaptation evolves; constituting an operational, though constrained, form of meta-recursion through open-ended search and selection.

\begin{quote}
\textit{Figure 4. [Image removed in preprint version]}\\
Work referenced: Conceptual diagram of the \textit{Darwin--Gödel Machine}, 2025. Schematic representation of a self-improving code agent combining a Gödel-style self-referential core with a Darwinian open-ended evolutionary search process. Diagram created by the author for scholarly illustration.
\end{quote}

\subsection{Synthesis}

\begin{table}[htbp]
\centering
\caption{Cross-Mapping Recursivity Levels, Operational Criteria ($\mu$/$\rho$/$R$), and Case Studies}
\label{tab:case-studies}
\small
\begin{tabularx}{\textwidth}{L{2.3cm} C{1.8cm} L{2cm} L{2cm} L{2cm} L{3cm}}
\toprule
\textbf{Case / System} & \textbf{Recursivity Level} & \textbf{State Memory ($\mu$)} & \textbf{Rule Evolution ($\rho$)} & \textbf{Reflexive Visibility ($R$)} & \textbf{Operational Characterization} \\
\midrule
Refik Anadol -- \textit{Machine Hallucinations} & 0 $\to$ 1 & Low (no internal state; optional external data) & None (fixed GAN; only parameter modulation) & Low (perceived continuity without structural change) & Iterative generation with aesthetic appearance of evolution \\
\addlinespace
Sougwen Chung -- \textit{Drawing Operations} & 2 & Medium--High (growing gesture database) & Medium (periodic retraining updates parameters) & Medium (feedback loop partly legible) & Co-evolving human--robot system \\
\addlinespace
Karl Sims -- \textit{Genetic Images} & 3 & High (heritability of code components) & High (mutations rewrite generative rules) & High (rule evolution visible through lineage) & Evolutionary system where rules evolve under fixed metarule \\
\addlinespace
Darwin--Gödel Machine & 4 (constrained) & Very high (entire codebase mutable) & Very high (self-modifying routines) & Very high (self-improvement explicit) & System altering principles of its own adaptation \\
\bottomrule
\end{tabularx}
\end{table}

\subsection{Concrete Pathways for Recursivism Without AI}

Recursion depends not on technology but on a rule applied to its own effects.

\textbf{Transformation protocol:} At each iteration, a parameter from the previous work is selected and transformed; crucially, the criteria for selecting and transforming parameters may themselves shift over time, based on the accumulated trajectory of the work.

\textbf{Social recursion:} Public participation does not merely generate variants but progressively reshapes the implicit norms governing what counts as a valid transformation.

These approaches show that Recursivism is above all a conceptual discipline: it formalizes the continuity of transformations in a world where everything becomes version, trace, or derivation.

\section{Aesthetic and Curatorial Implications}

\subsection{Aesthetic Evaluation: From Product to Protocol}

Recursivism shifts aesthetic evaluation from static objects to the evolving interplay between process and outcome, where each result manifests the system's transformation. The aesthetic experience is temporal and cumulative: it unfolds over observation duration, sometimes across multiple sessions.

New criteria emerge:
\begin{itemize}
    \item \textbf{Elegance of the recursive function:} clear, parsimonious, generative rule.
    \item \textbf{Depth of state memory:} effective integration of history, beyond superficial variations.
    \item \textbf{Richness of emergences:} unpredictable diversity without chaotic drift or mechanical repetition.
    \item \textbf{Conceptual readability:} perceptible or intelligibly documented loop.
\end{itemize}

These criteria extend conceptual art \cite{lewitt1967}, but here the ongoing protocol takes precedence over final result.

\subsection{Conservation and Exhibition}

Central challenge: preserve a work whose essence is continuous transformation while its supports are perishable \cite{rinehart2014, dekker2018}.

Possible curatorial approaches:
\begin{itemize}
    \item \textbf{Protocol documentation:} archive the recursive rule and its parameters (score, code, seeds, versions).
    \item \textbf{Selective freezing:} capture landmark states without interrupting the active loop.
    \item \textbf{Temporal window:} ``live'' exhibition of a recursion segment.
    \item \textbf{Meta-visualization:} display current state and a view of state space or history (logs, graphs).
\end{itemize}

These methods shift the museum from mausoleum of objects to laboratory of processes.

\subsection{Ethical Responsibility and Sustainable Public Participation}

When the loop incorporates external data or interactions, the public becomes system co-agent. Contributions leave persistent traces that influence future iterations.

Points of attention:
\begin{itemize}
    \item \textbf{Informed consent:} understanding of recursive usage and its duration.
    \item \textbf{Deferred agency:} potential long-term effects of initial gestures.
    \item \textbf{Data property and governance:} contributors' rights vs.\ protocol integrity.
\end{itemize}

Recursive transparency is desirable: provenance, trajectory and data reuse documented as much as possible \cite{gebru2021}.

\section{Critical Perspectives and Limitations}

While Recursivism offers a productive framework, several critiques warrant consideration:

\subsection{Risk of Technological Determinism}
Emphasizing AI's role risks implying that technology autonomously drives artistic evolution. This would invert actual causality: artists choose to engage AI precisely because it aligns with existing conceptual concerns. Recursivism must remain a chosen framework, not an inevitable consequence of available tools.

\subsection{Accessibility and Exclusion}
Constructing recursive AI systems requires significant technical expertise and computational resources. This potentially restricts Recursivism to well-resourced practitioners, raising questions about democratization of artistic means. However, this concern applies broadly to technology-intensive art forms and may be mitigated through open-source tools and collaborative infrastructures \cite{mansoux2008}. Moreover, the present framework is largely grounded in Euro-American art histories and AI infrastructures. Extending Recursivism will require engaging other artistic genealogies and cosmotechnical traditions, where recursion, memory, and self-modification may be conceptualized and practiced differently.

\subsection{Environmental Costs}
Training and operating AI models consume substantial energy. Recursive systems, continuously retraining, amplify this impact. Artists working within Recursivism bear responsibility for considering environmental sustainability, a dimension often overlooked in enthusiasm for technical possibility \cite{crawford2018}.

\subsection{Authorship and Authenticity}
When a system autonomously generates variations, traditional authorship concepts strain. Is the artist the system's architect, curator of its outputs, or collaborator with an emergent agent? These questions extend longstanding debates in generative and conceptual art but acquire new urgency when recursion produces outcomes unforeseen by the designer \cite{mccormack2019}.

\section{Conclusion}

Recursivism proposes to redefine art in the artificial intelligence era as reflexive function: a process that acts on its own effects and auto-transforms.

By revealing the impossibility of stable style or ultimate foundation, it shifts creation toward pure conceptual creativity: designing systems capable of self-modification.

Artificial intelligence appears to literalize this logic: learning, feedback, and retraining. The practices of Sougwen Chung and Karl Sims illustrate this passage: value resides not in formal execution but in designing the recursivist protocol.

Recursivism thus acts as interdisciplinary bridge between art, science, and technology, providing common language to think transformation, memory, and emergence across self-evolutionary systems.

It offers artists conceptual control over their creative loops: an architectural capacity rather than mere machine use. Art becomes a field of aesthetic engineering: not production of finished objects but conscious orchestration of reflexive processes where our modes of technical and cultural evolution replay themselves.



\begin{thebibliography}{99}

\bibitem{ivins1953}
W.~M. Ivins Jr., \textit{Prints and Visual Communication} (Cambridge, MA: Harvard University Press, 1953).

\bibitem{scharf1968}
Aaron Scharf, \textit{Art and Photography} (London: Allen Lane, 1968).

\bibitem{paul2015}
Christiane Paul, \textit{Digital Art}, rev.\ ed.\ (London: Thames \& Hudson, 2015).

\bibitem{graham1994}
Ronald~L. Graham, Donald~E. Knuth, and Oren Patashnik, \textit{Concrete Mathematics: A Foundation for Computer Science}, 2nd ed.\ (Reading, MA: Addison-Wesley, 1994).

\bibitem{hofstadter1979}
Douglas~R. Hofstadter, \textit{Gödel, Escher, Bach: An Eternal Golden Braid} (New York: Basic Books, 1979).

\bibitem{darwin1859}
Charles Darwin, \textit{The Origin of Species} (1859; repr., New York: Signet Classics, 2003).

\bibitem{maturana1980}
Humberto~R. Maturana and Francisco~J. Varela, \textit{Autopoiesis and Cognition: The Realization of the Living} (Dordrecht: D.\ Reidel, 1980).

\bibitem{baudelaire1965}
Charles Baudelaire, ``The Modern Public and Photography,'' in \textit{Art in Paris 1845--1862: Salons and Other Exhibitions Reviewed by Charles Baudelaire}, trans.\ and ed.\ Jonathan Mayne (London: Phaidon Press, 1965).

\bibitem{malevich1990}
Kazimir Malevich, ``From Cubism and Futurism to Suprematism: The New Painterly Realism'' (1915), in \textit{Malevich: Artist and Theoretician}, trans.\ Sharon McKee (Paris: Flammarion, 1990).

\bibitem{reinhardt1991}
Ad Reinhardt, \textit{Ad Reinhardt} (New York: Rizzoli, 1991).

\bibitem{derrida1976}
Jacques Derrida, \textit{Of Grammatology} (1967), trans.\ Gayatri Chakravorty Spivak (Baltimore: Johns Hopkins University Press, 1976).

\bibitem{shannon1948}
Claude~E. Shannon, ``A Mathematical Theory of Communication,'' \textit{Bell System Technical Journal} 27, no.\ 3 (1948): 379--423.

\bibitem{russell2021}
Stuart~J. Russell and Peter Norvig, \textit{Artificial Intelligence: A Modern Approach}, 4th ed.\ (Harlow, UK: Pearson, 2021).

\bibitem{galanter2003}
Philip Galanter, ``What Is Generative Art? Complexity Theory as a Context for Art Theory,'' in \textit{GA2003 -- 6th Generative Art Conference} (Milan, 2003).

\bibitem{boden2009}
Margaret~A. Boden and Ernest~A. Edmonds, ``What Is Generative Art?,'' \textit{Digital Creativity} 20, nos.\ 1--2 (2009): 21--46.

\bibitem{wiener1948}
Norbert Wiener, \textit{Cybernetics: Or Control and Communication in the Animal and the Machine} (Cambridge, MA: MIT Press, 1948).

\bibitem{vonfoerster1974}
Heinz von Foerster, \textit{Cybernetics of Cybernetics} (Urbana: University of Illinois Biological Computer Laboratory, 1974).

\bibitem{morris1969}
Robert Morris, ``Continuous Project Altered Daily,'' \textit{Artforum} 7, no.\ 8 (April 1969): 50--54.

\bibitem{cornock1973}
Stroud Cornock and Ernest~A. Edmonds, ``The Creative Process Where the Artist Is Amplified or Superseded by the Computer,'' \textit{Leonardo} 6, no.\ 1 (1973): 11--16.

\bibitem{whitelaw2004}
Mitchell Whitelaw, \textit{Metacreation: Art and Artificial Life} (Cambridge, MA: MIT Press, 2004).

\bibitem{anadol2021}
Refik Anadol, \textit{Machine Hallucinations: Nature Dreams} (AI data sculpture, 2021), König Galerie. Online: \url{https://refikanadol.com/works/machine-hallucinations-nature-dreams/}

\bibitem{chung2018}
Sougwen Chung, \textit{Drawing Operations} (artist project), ongoing human--machine collaborative drawing series. Online: \url{https://sougwen.com/work/drawingoperations2018/}

\bibitem{sims1991}
Karl Sims, ``Artificial Evolution for Computer Graphics,'' in \textit{SIGGRAPH '91 Proceedings} (New York: ACM, 1991). Project documentation: \url{https://karlsims.com/genetic-images.html}

\bibitem{zhang2025}
Jenny Zhang, Shengran Hu, Cong Lu, Robert Lange, and Jeff Clune, ``Darwin Gödel Machine: Open-Ended Evolution of Self-Improving Agents,'' arXiv preprint arXiv:2505.22954 (2025). \url{https://arxiv.org/abs/2505.22954}

\bibitem{lewitt1967}
Sol LeWitt, ``Paragraphs on Conceptual Art,'' \textit{Artforum} 5, no.\ 10 (1967): 79--83.

\bibitem{rinehart2014}
Richard Rinehart and Jon Ippolito, \textit{Re-collection: Art, New Media, and Social Memory} (Cambridge, MA: MIT Press, 2014).

\bibitem{dekker2018}
Annet Dekker, ed., \textit{Collecting and Conserving Net Art: Moving Beyond Conventional Methods} (London: Routledge, 2018).

\bibitem{gebru2021}
Timnit Gebru et al., ``Datasheets for Datasets,'' \textit{Communications of the ACM} 64, no.\ 12 (2021): 86--92.

\bibitem{mansoux2008}
Aymeric Mansoux and Marloes de Valk, eds., \textit{FLOSS+Art} (London: OpenMute / GOTO10, 2008).

\bibitem{crawford2018}
Kate Crawford and Vladan Joler, \textit{Anatomy of an AI System: The Amazon Echo as an Anatomical Map of Human Labor, Data and Planetary Resources} (New York: AI Now Institute, 2018).

\bibitem{mccormack2019}
Jon McCormack, Toby Gifford, and Patrick Hutchings, ``Autonomy, Authenticity, Authorship and Intention in Computer Generated Art,'' in \textit{Computational Intelligence in Music, Sound, Art and Design}, ed.\ Tiago Martins et al.\ (Cham, Switzerland: Springer, 2019).

\end{thebibliography}
\end{document}